# Towards unified Geophysical Data Requirements for Magnetic Navigation (MagNav)

Regupathi Angappan, Kimberly Moore, Sriharsha Thoram*
SandboxAQ, Palo Alto, CA, USA

*Corresponding author: harsha.thoram@sandboxquantum.com

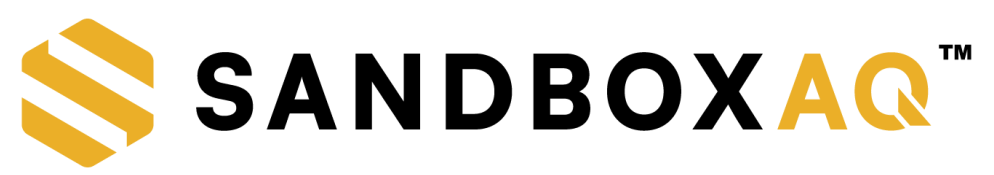

## Executive Summary

The United States Government has identified the urgent need for resilient alternatives to GPS/GNSS to support national Positioning, Navigation, and Timing (PNT) capabilities. Among the most promising of these alternatives is **Magnetic Navigation (MagNav)**, which leverages the Earth's magnetic field as a naturally stable and globally available reference for absolute positioning.

A critical enabler of MagNav is the **availability and quality of magnetic anomaly maps**, a specialized geospatial dataset derived from measurements of the Earth's crustal magnetic field. The successful development, deployment, and global scalability of MagNav depend heavily on improving the resolution, consistency, accessibility, and error quantification of these datasets. In prior efforts, the United States and allied governments have formalized geophysical model standards—such as the **World Magnetic Model (WMM)**—through clearly defined Requirements.

This white paper initiates a community-focused dialogue on **future geophysical data needs for MagNav**, grounded in our operational experience. At SandboxAQ, we have conducted extensive MagNav flight trials under real-world conditions, systematically evaluating the performance of existing geophysical data products.

We present two primary MagNav use cases with distinct data needs:

- **Operational MagNav**, requiring robust, globally consistent, high-fidelity datasets for navigation deployment.

- **MagNav Research and Development (R&D)**, requiring flexible, high-resolution data with transparent metadata and provenance to support innovation.

For each use case, we offer a prioritized set of recommendations for future geophysical data Requirements, along with justifications based on our empirical findings. We also identify specific areas for near-term investment, such as **designated test ranges** and **expanded magnetic anomaly coverage**.

Finally, we propose a key enhancement: **extending the World Magnetic Model (or equivalent models) to spherical harmonic degree 13**, aligning it with the resolution commonly used in Magnetic Anomaly Field modeling. This upgrade would improve compatibility across MagNav systems and support more accurate, seamless integration into alt-PNT architectures.

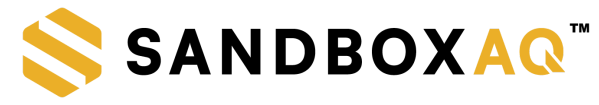

## TABLE OF CONTENTS

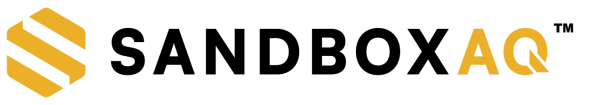

## 1. Introduction

**Magnetism has formed the basis of navigation for millenia.** Early compasses allowed for navigation, by providing the overall direction of Earth's magnetic field. Today, modern Magnetic Navigation (MagNav) goes leaps beyond, using precision quantum sensing measurements to provide exact quantitative values of the Earth's magnetic field globally. These measurements are commonly combined with an inertial navigation system as part of a sensor fusion solution.

**Magnetic Navigation (MagNav) is a critical technology that can provide an alternative Position, Navigation, and Timing (alt-PNT) solution when GPS/GNSS is degraded or denied.** GPS and GNSS form the basis of modern navigation, but are not consistently available worldwide. For example, GPS/GNSS signals are not available underwater, a key navigational use case. But even when GPS/GNSS referencing is possible, the signal may be disrupted by methods such as spoofing and jamming. These are common techniques to degrade and disrupt access to critical PNT information. They impact both commercial and defense users alike, from passengers to global shipping and operations. For example, using data from FlightRadar24, the Financial Times reported in May 2024 that over 40% of air traffic is routinely jammed in the capital city of Ankara, Turkey and other major international cities (Murray et al.). Overall prevalence of jamming/ spoofing is also increasing. In Figure 1, we highlight the daily prevalence of GPS Spoofing/Jamming compiled by GPSJAM.ORG; significant outages are visible near conflict zones such as Ukraine or Gaza. *For a broader review of GPS/GNSS vulnerabilities and alt-PNT technology requirements, see, for example, "Report on Positioning, Navigation, and Timing (PNT) Backup and Complementary Capabilities to the Global Positioning System (GPS) (Dept. of Homeland Security, 2020).*

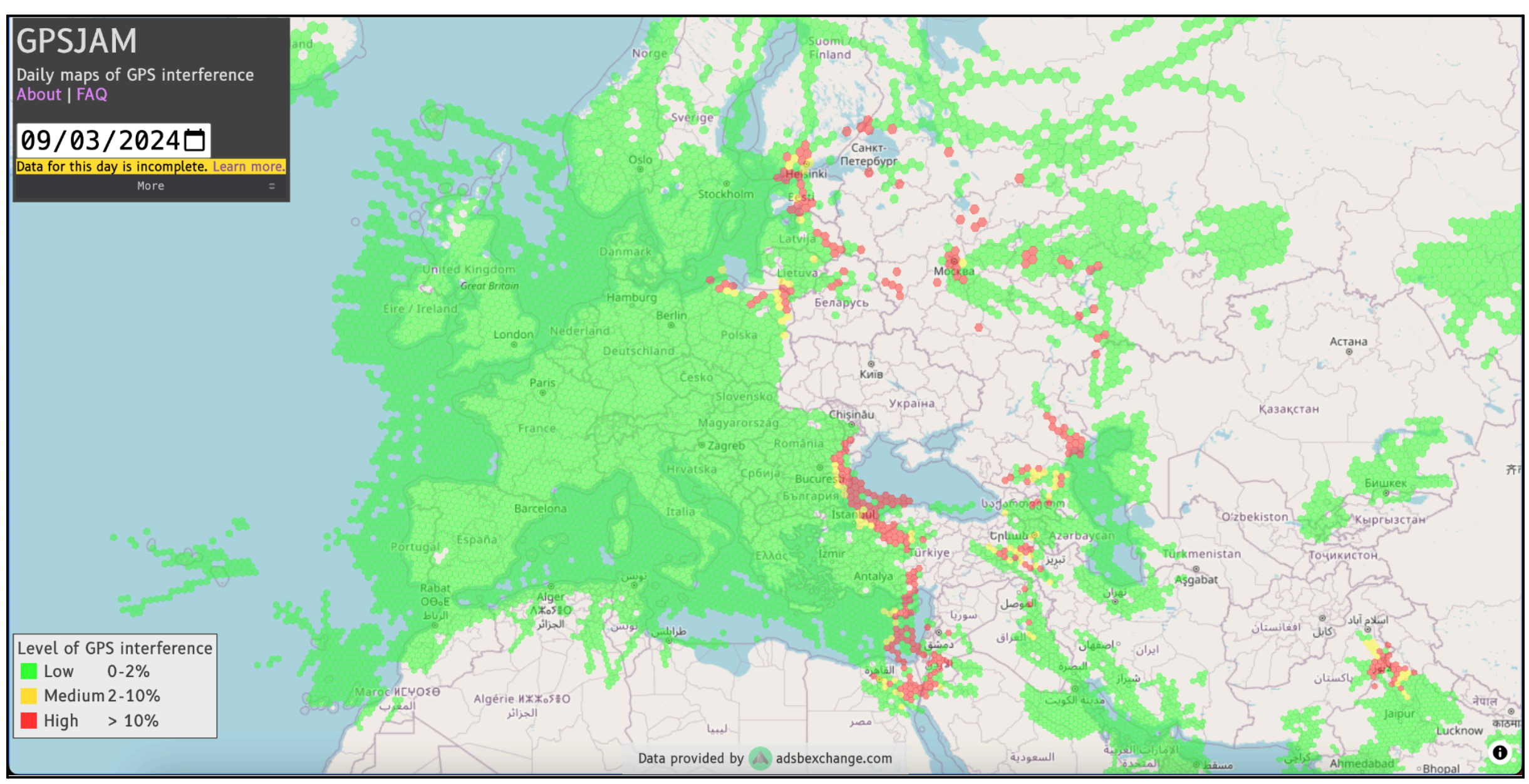

**Figure 1: Example illustration of GPS Jamming prevalence for September 3rd, 2024.**
Source: GPSJAM.org, by John Wiseman. Screenshot.

**Alt-PNT techniques like magnetic navigation can serve as reliable backups or complements to GPS, fulfilling critical infrastructure needs. To date, extensive MagNav test flights have shown sufficient accuracy to successfully enable key commercial and defense alt-PNT use cases.** Early studies demonstrated high performance in post-processing. For example, 2016 cross-country test-flights showed positional accuracy of 3.2 km over the course of a 5 hour flight[1] (Canciani, 2016). Further testing aboard a F-16 aircraft with USAF test pilots demonstrated post-processing positioning positional accuracy of ~50 m over a 1.5 hr flight[2] (Canciani, 2021). In both cases, the MagNav systems beat the performance of an unaided INS by over a factor of 10.

Building on a series of rapid advancements, SandboxAQ continues to redefine the capabilities of Magnetic Navigation (MagNav) across diverse platforms. In July 2024, the team demonstrated real-time MagNav aboard a U.S. Air Force C-17 at Joint Base Charleston, marking the first time MagNav served as the primary method of navigation in flight—a milestone described as a ‘huge step’ toward providing Assured PNT in contested environments. More recently, in early 2025, SandboxAQ achieved a critical certification milestone by receiving flight approval from the C-130J Systems Program Office (SPO). This cleared the way for participation in Exercise Emerald Flag, where the system successfully completed two oceanic flights in the Gulf of Mexico, totaling over eight flight hours of resilient performance.

Beyond domestic military exercises, SandboxAQ is also scaling this technology through a strategic collaboration with Airbus’s Acubed. Following successful flight trials on a Beechcraft Baron, the partnership focused on maturing ‘MagNav at scale’—integrating high-performance quantum sensors and AI-driven magnetic interference compensation to a flight campaign of 150 hours, 44,000 kilometers of flight profiles (Blog).

**Broadly, these flight trials over the past decade demonstrate that the time for MagNav is now – that the technology is sufficiently mature to consider the strategic implications for key supporting geomagnetic data products.** Overall, it is critical to enable alt-PNT technologies such as MagNav, in order to provide robust PNT solutions for future commercial and defense navigation. In the next sections, we take a deeper look at the geomagnetic data underpinning this important technology.

---

[1] Specifications: *Tactical-grade INS, 2.7 km altitude above ground level, tail stinger -mounted magnetometer, low/medium spatial resolution magnetic anomaly map (NAMAM over the midwestern United States)*

[2] Specifications: *Navigation-grade INS, 300 m altitude above ground level, sensor placement in a RASCAL pod configuration, high-quality modern magnetic survey map (Edwards Air Force Base / Sanders Geophysics Limited)*
*A second flight at 1.5 km altitude yielded a positional accuracy of 110 m in post processing.*

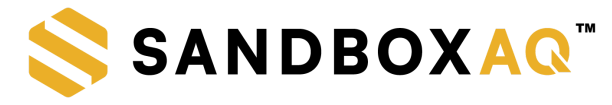

## 1.1 Maps for MagNav

**Geomagnetic data is an integral part of MagNav.** Specifically, to provide a PNT solution, a MagNav system compares input from one or more magnetic field sensors to a reference field. This reference typically encompasses a core field model, and a crustal field model (i.e. a magnetic anomaly field map). Depending on accuracy requirements, additional field sources may also be considered, such as Space Weather (see Angappan et al., 2024). *This present white paper will focus primarily on MagNav data requirements for crustal field and core field modeling.*

Currently, there are many publicly available magnetic anomaly datasets for the United States and beyond. For example, the National Oceanic and Atmospheric Administration - National Centers for Environmental Information (NOAA NCEI) maintains a Marine Trackline Geophysical Database, an archive of geomagnetic and other geophysical data from individual marine and aeromagnetic surveys of the ocean. The United States Geological Survey (USGS) maintains at least two separate databases of individual magnetic anomaly surveys: a baseline database containing historical surveys by US State and County (Bankey et al., 2002), as well as the Earth Mapping Resources Initiative (Earth MRI) database of recent high-resolution surveys. Several U.S. government agencies also maintain larger regional/continental models, such as NAMAM (Bankey et al., 2002), NAMAM NURE (Ravat et al., 2009); and global models, such as EMAG2v3 (Meyer et al., 2017), and the High Definition Geomagnetic Model series (HDGM, HDGM-RT; see NOAA, Nair et al. (2021)). Finally, beyond the United States, local surveys are available through other governments or academic collaborations. For example, the World Digital Magnetic Anomaly Map (WDMAM; Choi et al.) is maintained jointly through the International Association of Geomagnetism and Aeronomy (IAGA) and the Commission for the Geological Map of the World (CGMW).

**But while many geomagnetic anomaly datasets are publicly available, they are not always readily usable for MagNav technology development and deployment**. Existing datasets have largely been shaped by requirements from other communities. In particular, this includes fundamental science investigations (academic research), natural hazards forecasting, and the Natural Resources Industry. For example, the NOAA HDGM model, while publicly available to all, is primarily licensed by major natural resources companies, particularly for drilling use cases (as of 2024, license holders included Schlumberger, Halliburton, Scientific Drilling, and more; see full list on the NOAA HDGM website). Magnetic Navigation use cases may have unique needs for geophysical data products, such as spatial resolution, error estimation, and scalability. We detail our perspective on these needs further in Section 3, to drive discussion within the geophysics and navigation industry communities.

## 1.2. Standardization of Geophysical Data for MagNav

**Geospatial data used for Government or Defense use cases often goes through a Requirements and standardization process.**

For example, the World Magnetic Model (WMM) is a standard model of Earth's core field, periodically published every 5 years, with occasional mid-term updates as needed (Chulliat et al., 2020, 2019). It is owned by the United States National Geospatial-Intelligence Agency (NGA) and the United Kingdom's Defence Geographic Centre (DGC), and developed in partnership with other agencies and data providers. Specifically, the WMM *"the standard [core field] model for navigation, attitude, and heading referencing systems using the geomagnetic field"* (NOAA - NCEI).

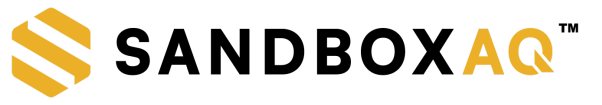

**The WMM geophysical model in its current form is based on Requirements, outlining the proposed specifications and needs of the Defense and Intelligence communities.** [Unclassified] Military Specification: World Magnetic Model (MIL-W-89500) was first published in 1993, which states: "*The purpose of this document is to specify the WMM and the Defense Mapping Agency (DMA) products that are derived from the model. This document also specifies the accuracy and limitations of the model."* These requirements were later updated with MIL-PRF-89500A (2015) and B (2019), which added additional performance constraints based on need. Similar requirements were adopted globally, such as by the North Atlantic Treaty Organization (NATO), e.g. NATO STANAG 7172, "Use of Geomagnetic Models" (2017). Prior to these requirements, earlier versions of the WMM (WMM 1990-1995) were produced by the US Naval Oceanographic Office and the British Geological Survey (Quinn et al., 1991).

**Magnetic anomaly Navigation is a critical Defense and navigational use case, with significantly different data needs than existing Geomagnetic Model requirements.** While this Paper will focus on potential data priorities for Magnetic Navigation exclusively, we note that other forms of geophysical navigation (such as gravitational or terrain-based systems) may share a substantial overlap in technical data needs. For example, Crustal magnetic field datasets and distribution formats may share more in common with other non-spatial geophysical data (e.g. terrain models), than with standard spherical harmonic core field models such as the WMM.

We provide recommendations for MagNav use cases, and their corresponding data needs in the following Sections.

## 1.3 White Paper Overview

**In this White Paper, we provide recommendations for future magnetic anomaly and core field model Requirements, based on our experience with key MagNav Use Cases.** Our goal is to promote discussion within the navigation industry and geophysics communities, and to enable our government partners as they evaluate the geospatial data needs of tomorrow. Leveraging our years of experience working with geophysical datasets for Magnetic Navigation in defense and commercial settings, we provide several recommendations in the following sections.

*Note: This White Paper focuses on the potential MagNav data requirements for crustal magnetic field (i.e. magnetic anomaly) and core field models. We will primarily detail recommendations for scalar magnetic anomaly data, but provide some limited context on vector anomaly data as well.*

## 1.4 Scope & Assumptions

The recommendations in this White Paper are framed around magnetic navigation systems intended for integration with inertial navigation systems (INS), rather than as standalone navigation solutions. Unless otherwise specified, we assume the use of scalar magnetic anomaly measurements as the baseline operational modality, recognizing that vector anomaly data represents a promising but less mature capability primarily relevant to future research and test-range development. The discussion is oriented toward airborne and maritime platforms, including both crewed and uncrewed systems, and assumes navigation-grade or tactical-grade sensors typical of current MagNav deployments. These assumptions are made to focus the discussion on broadly applicable geophysical data requirements, rather than platform-specific implementation details.

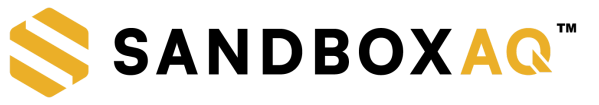

## 2. MagNav Community Use Cases

In this Section, we outline two primary MagNav community use cases, in order to contextualize the standardization of geomagnetic maps for MagNav. Specifically, we discuss:

1. Operational MagNav (Global, on-device deployment) - good map coverage
2. MagNav Research & Development (R&D)

We believe that the final requirements for map datasets in these use cases will be distinct. However, we find that in community dialogue, these use cases are mixed interchangeably; they should not be. We would like to clarify and distinguish the individual use cases to guide future discussions and map products. In Section 3, we will dive into the specific map data recommendations for each of these high-level use cases, based on definitions in Section 2. *Each use case may have subcategories depending on industry/mode of transportation (e.g. aerospace and maritime), with more tailored technology requirements. However, we will focus only on the high-level needs here.*

For each use case, we name 1) the Community driving the use case (i.e. who will be using the geophysical data directly), as well as 2) the intended Data Users of the standardized geophysical data. *We note that the Data Users may be distinct from the overall MagNav Product End-Users. For example, a commercial company could be considered a geophysical data user, who ingests geophysical data and integrates it into their navigation product. But an individual pilot deploying MagNav technology on their aircraft may only be an End-User, and would only view the overall navigation solution, without a direct need to access the intermediate geomagnetic data itself. This is comparable to viewing raw INS data (gyro, acceleration, etc) versus an End-Use INS-only Navigation Solution.*

### 2.1 Operational MagNav deployment (On-Device)

**This is the primary use-case typically associated with MagNav: a working magnetic navigation product deployed in the field,** which enables navigation integrity and reliability for real users. For this use case, we have the following definitions:

- **Community driving the use case:** commercial industry or federal transition partners
- **Geophysical Data Users**: same as above; the entities building the products
- **MagNav Product End-Users:** individual users seeking to deploy the technology in the field (pilots, captains, commanders, etc)

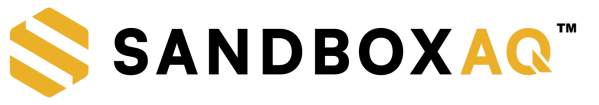

In this use-case, a MagNav product solution/system is deployed on one or more operational platforms (e.g. one or more vehicles), in commercial and/or defense settings. The overall product deployment will likely have stringent Size, Weight, and Power and Compute (SWaP-C) requirements. It may also be subject to stringent reliability and certification requirements. Like other kinds of non-visual data for alt-PNT technology (e.g. gravity, terrain, etc), Data Users may wish to view the data for software testing and validation, but End-Users will generally have no need to see original source data during Operational Product use.

Magnetic field datasets used in this use case will likely have to meet requirements for spatial coverage, uncertainty knowledge, and scalability. It will also be important to have a single, "unified/merged map" for use in deployment, to ensure a map is validated and computationally inexpensive to use. We discuss this idea in more detail in Section 3.

## 2.2 MagNav R&D

**This secondary use case is also essential to community development of MagNav Technology. Specifically, MagNav R&D includes all stages of the development pipeline:** from *in silico* simulations of algorithms, to test flights of new MagNav prototype versions, to validation of existing performance. MagNav R&D may even include purely geophysics-focused R&D.

For this use case, we have the following definitions:

- **Community driving the use case:** academia, research labs, and commercial industry
- **Geophysical Data Users**: Same as above
- **MagNav Product Users:** Same as above; R&D for internal or external use

Unlike Operational MagNav, for this use case the MagNav community will wish to have full access to all aspects of the original dataset(s). Data users will wish to understand the detailed interactions between MagNav algorithms/performance and the underlying map datasets, under a variety of different conditions and assumptions. Metadata, raw and merged data, processing reports/history, etc. could all play a role. Data Users will also desire expanded data visualization capabilities, including potentially searchable databases and/or GUIs.

## 2.3 A Note on Sparse Maps

We would like to highlight the situation where map data may not be available, or have limited resolution. As discussed, map coverage of the Earth is not geographically uniform. Some areas of the Earth have substantially higher resolution geomagnetic anomaly map coverage than others. As an example of the broad range of resolutions to a single anomaly map product, input datasets to the EMAG2v3 model range in spatial resolution from 1 km to 20 km grid spacing (Meyer et al., 2017).

In the long term, the ideal state is to have perfect map coverage everywhere – but depending on specific technology and navigation requirements *(e.g. RNP10 vs RNP1 vs 1 m accuracy)*, map resolution may not always be sufficient.

The MagNav performance of individual map datasets can be simulated (Bell, 2024), to guide recommendations for their approved use. However, that does not change the fact that in emergency scenarios, alt-PNT will still be desired in regions with lower map availability. This suggests a thoughtful approach to enabling both R&D and deployment to areas with sparse map coverage. This could potentially be treated as a separate use case, with different requirements. It is both an opportunity for technological innovation and operational enablement, as well as a key area to highlight for further strategic consideration.

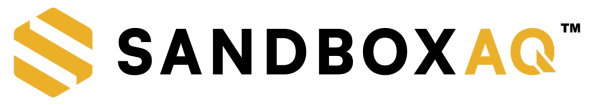

## 3. Suggested Geophysical Data Requirement Areas by Use Case

In this Section, we suggest focus areas for formal Requirement development for the two main MagNav use cases identified in Section 2: Operational MagNav Deployment (on-device), and MagNav R&D. **We prioritize fundamental, high-level needs that will be true for the majority of industry, government, and academic institutions**. **We recommend that the navigational and geophysics communities focus on high-priority, top-level needs first**, before engaging in debates over finer details that are likely user-specific. Otherwise, major top-level needs may not be discovered and agreed upon until too late a time in the requirements process.

For example, we consider coordinate information to be a fundamental need for MagNav data products. All users will require data to have coordinate information in some form. But at the finer levels of detail, each user may have user-specific preferences for the specific formatting, e.g.:

- **Company A** has navigational code that uses data formatted as Nx3 arrays in the order of ['LAT', 'LON', 'ALT'], in units of deg:mm:ss and meters, provided to 8 decimal places
- **Government Institution B** uses a data format of ['LON', 'LAT', 'ALT'] in 3xN arrays, in units of radians and meters, to double precision
- **Research Lab C** prefers a third format, etc.

We seek to avoid this situation. We do not dismiss the importance of user-specific preferences to an individual user in the MagNav community; indeed, these preferences can be deeply meaningful in terms of resources required (cost, time, and staffing). Converting a massive geophysical data product from one format or coding language to another can be time-consuming and costly, and may require detailed investigations of legacy code. However, while a top-level requirement like "a unified coordinate system for map data" affects all users, community requests at finer levels of detail will still be specific to just individual users, and should be a lower priority in discussion across the whole community. Ultimately, the final, overall map requirements should still be designed such that all users are able to use the final product, even if it is more convenient for some than others.

Below, we provide an overview of our identified priority areas for each use case, from our perspective as a geophysical dataset user for MagNav. We also provide tables summarizing key discussion areas for the community, with our corresponding recommendations.

### 3.1 Interpretation of Recommendations

The priorities outlined in this section are intended to guide future requirements development rather than prescribe fixed specifications. We distinguish between foundational capabilities necessary for any credible MagNav deployment, high-priority enhancements that substantially improve operational robustness, and longer-term opportunities that enable future innovation. This structure is intended to support phased implementation, allowing the community to focus first on universally enabling requirements before addressing more specialized or platform-dependent considerations.

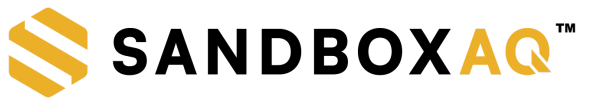

### 3.2 Operational Magnetic Navigation

The Operational MagNav use case relies on a single cohesive (merged), queryable dataset that provides magnetic anomaly magnitude and uncertainty values at maximum resolution across the globe. In particular, we envision the following properties as essential – described below, and summarized in Table 1.

**Table 1. Priorities for Geophysical Datasets – Operational MagNav Use Case**

| Dataset Property | Recommendation |
|---|---|
| Dataset Cohesion | Global, merged dataset *(not just individual survey files)* |
| Coordinate System | Uniform coordinate system *(Standardized & Known)* |
| Geophysical Data Fields | Latitude, Longitude, Altitude, Magnetic Anomaly Field |
| Magnetic Field Uncertainty Estimates (3D) | Localized, queryable uncertainty estimates in 3D space *(May be provided as separate dataset/model, or as an additional entry in the above)* |
| Spatial Resolution | Maximum resolution *(non-uniform OK)* |
| Altitude | Lowest altitude |
| Processing | Standardized processing applied to all datasets (when possible)<br><br>Prioritize true representation of the data<br>● Not reduced to pole<br>● Consistent, thoughtful approach to geophysical signals from infrastructure (“cultural artifacts”) |
| File Format | Non-proprietary data format |
| Queryability | A baseline map/model that is designed to be queryable.<br><br>There are multiple options, any of which can serve the community need e.g.<br>a) A single map layer at the lowest altitude, which users can upwards continued to the desired points<br>b) An equivalent source model that allows users to easily interpolate the field anywhere |

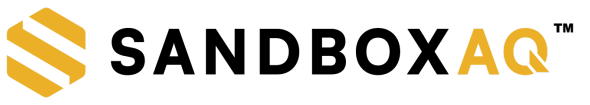

| | |
|---|---|
| | Etc. |
| Customizable Scalability | Queryable dataset/model must be scalable to small SWaP-C systems; dataset must be able to be easily subdivided into smaller files by the users.<br><br>Source data should be accessible, and not locked behind a rigid API. |
| Documentation | Clear documentation describing processing applied to the dataset |
| Updates | Systematic cadence of dataset updates |

**Geophysical Data Fields & Coordinate System:** At the most basic level, Operational MagNav users will require a magnetic anomaly dataset consisting of latitude, longitude, altitude, and magnetic field magnitude (*We discuss vector magnetic maps in Section 4, regarding MagNav R&D Test Ranges)*. We recommend providing all magnetic field data in a standardized coordinate system, ideally one also common in navigational use cases. Current public magnetic field datasets are in a wide range of coordinate systems; some of these are commonly used *(e.g. Emag2V3 uses WGS-84 coordinate system; Meyer et al., 2017)*, while others are published in uncommon systems or do not have well-publicized transforms. We also highly recommend adopting a coordinate system common such as WGS84 (the standard for global navigation; NGA Office of Geomatics) or EGM2008/Mean Sea Level (MSL). If a coordinate system that is referenced to terrain is used (e.g. Above Ground Level (AGL)), we suggest a standardized Digital Elevation Model (DEM), like the Digital Terrain Elevation Data (DTED), also be specified, so that all users can perform the same transform and get the same results when the map is queried.

**Geophysical Uncertainty Estimates (3D):** A potentially unique and critical requirement for Operational MagNav is full, queryable 3D magnetic field uncertainty estimates. Uncertainty modeling and evolution is an essential aspect of all Position, Navigation, and Timing (PNT) estimation. For example, in standard navigational Kalman filters, uncertainty estimates are incorporated directly into reference measurement models (for a broad review on State Estimation techniques, see Crassidis & Junkins, 2011). This lets the navigational filter know how much to trust any given magnetic field measurement (i.e., how to propagate the PNT state estimate), and how to weigh the magnetic field measurements compared to any other sensor systems which may be fused into the filter (i.e. sensor fusion). Current magnetic anomaly maps, particularly merged multi-survey models, are not always accompanied by uncertainty estimates. Further, models that do include errors often are highly conservative, to the extent that error estimates may be larger than the value of the field itself; e.g., in EMAG2v3, where the maximum uncertainty tier for non-error-coded data is 232 nT (Meyer et al., 2017).

We recognize that while important, uncertainty modeling is still a dynamic and evolving field in geophysical data analysis. Calculating or updating uncertainty estimates is a time-consuming and complex process. This is especially true when considering the range of effects that may contribute to uncertainty estimates: original survey processing methods (e.g. was solar weather removed?), data spikes or other artifacts, merging effects, uncertainty in regions where data gaps were in-filled, etc. Full documentation of original map processing methods may not be widely unavailable or even recorded, especially given many surveys were recorded over 50 years prior to today. However, exciting work is happening in this domain (e.g. Saltus, 2023). We advocate that error modeling for magnetic field data is of high priority to the MagNav community, and should be provided as it becomes available, with updates in ongoing geophysical data versions.

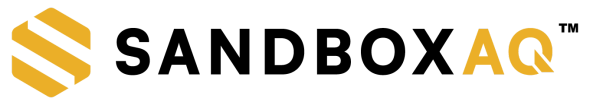

A second unique perspective is that MagNav will ultimately utilize full 3D uncertainty estimates. Each time a platform utilizing MagNav travels, it will perform Kalman filter updates by measuring the local magnetic field, denoising it as necessary, and then comparing it to a reference magnetic field map at the user's exact predicted location. As described above, uncertainty modeling is a key aspect of the Kalman filter update step calculation. This update process happens at all altitudes, and requires a local uncertainty comparison – i.e, for an aircraft flying at a cruising altitude of 20km, the uncertainty estimates for the base grid at sea level may not be sufficient. We recommend that either full 3D uncertainty estimates be provided. This could take the form of a direct method / model / API provided to the user for this extrapolation. Alternatively, it could take the form of extended technical guidance for the user on how to implement this calculation themselves.

**Dataset Cohesion (Single Merged Dataset):** Typically, magnetic field datasets are often been provided as local surveys *(e.g. USGS surveys (Bankey et al., 2002; Earth MRI))*, Marine tracklines *(e.g. NOAA NCEI marine trackline database)*, or regional continental grids *(e.g. NAMAM NURE, Ravat et al., 2009)*. However, we recognize that the mission requirements of Operational MagNav product users are unlikely to fit nicely within the irregular bounds of a pre-existing survey, or align perfectly along a single marine track line. For example, aviators frequently perform transoceanic flight, which would require crossing several local/regional boundaries throughout the duration of flight.

We believe having a cohesively merged map is a high priority for the community. From a technical perspective, merging surveys is a time-consuming and often manual process requiring extensive expertise to perform, and so cannot easily be done in real time/ on-device during operational MagNav usage. Further, individual surveys may also have extensive historical processing, require inside knowledge of artifacts, or utilize obscure coordinate systems. The process of treating long-wavelength signals in global maps is also non-trivial (e.g. Meyer et al., 2017; Choi et al.)

Thus, there is a benefit to having a single, cohesive, pre-merged global map available to all users. We note that some cohesive, global datasets do exist: for example, the WDMAM (Choi et al.), the EMAG-series (NOAA/NCEI; Meyer et al., 2017), and the NOAA HDGM model (NOAA NCEI; Nair et al., 2021). However, existing global datasets often downsample data to lower resolutions than may be required by Operational MagNav users, even for areas where higher resolution is available (e.g. EMAG2v3 vs NAMAM source grid resolution for the continental United States). We discuss this in the following item.

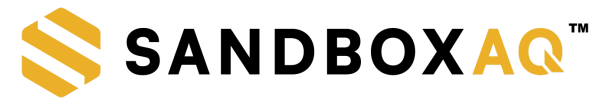

**Maximum Spatial Resolution:** We strongly advocate for the Operational MagNav geophysical dataset to be available in the highest possible spatial resolution. This could either be the resolution at which a map is considered “fully resolved” (fully sampled) at a given altitude, or the maximum resolution available if the map is considered under-resolved.

Importantly, we advocate not to over-smooth higher-resolution regions of data for the sake of achieving the lowest common denominator resolution. Rather, we advocate for gridding/modeling the data to the highest possible resolution, and simply design the dataset/API to be loaded in a scalable way (discussed more below in the next sections). Users can always easily downsample a map to lower resolutions if needed, but smaller wavelength signals cannot easily be added back to a map once they have been filtered or smoothed.

We note that non-uniform map resolution may be possible, or even encouraged. Many geophysical field calculation methods do not impose any requirements for uniform grids or resolution (e.g. equivalent source). Even for interpolation methods that do require uniform gridding along a given axis (e.g. Fourier), users implementing those methods could simply interpolate the low-resolution data onto a more finely spaced grid first. Such a user re-gridding would not require any new data, nor would it change the underlying uncertainty assigned to those grid points.

Ultimately, the overall map resolution required will depend upon the Operational MagNav mission profiles / platforms the community wishes to support. From a technical standpoint, the key metric to consider is the Required Navigation Performance (RNP; or international equivalents) of the traditional Navigation or MagNav system. Existing needs vary significantly across potential mission profiles; for example, some mission profiles / geographic regions may require RNP10, while others require performance equal to or better than RNP1. Until the desired RNP level for MagNav is firmly established, we prioritize ensuring that the full/highest possible spatial resolution data is available. Please *see also our upcoming SandboxAQ publication connecting map spatial resolution with upper bounds on MagNav performance (ref).*

**Altitude:** Adequate map coverage as a function of altitude will be essential to the Operational MagNav use case. MagNav will be desired for a full range of mission altitudes; many use cases require ground/sea-level maps, or even undersea maps. Geophysically, altitude often comes into play via upwards/downwards continuation. The Downward Continuation challenge is well-known in the geophysical community, wherein extrapolating data to lower altitudes is a mathematically unstable process, often requiring manual application of techniques such as regularization (e.g. Shure et al., 1982). This means that baseline data will be more useful to Operational MagNav users when it is provided at the lowest possible altitude(s), since users may not have the ability to downwards continue the data themselves.

In addition to providing magnetic anomaly values across altitude, Operational MagNav datasets would benefit from explicit specification of altitude validity bounds. These bounds define the vertical region over which the map predictions and associated uncertainty estimates are considered reliable, accounting for limitations introduced by upward or downward continuation and data sparsity. Providing altitude-dependent validity metadata would enable navigation systems to appropriately gate measurements, adjust uncertainty models, or gracefully degrade performance when operating outside trusted regions, thereby improving robustness and supporting certification and assurance processes.

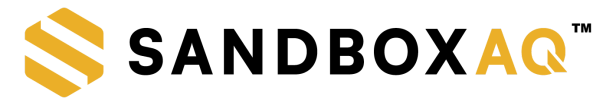

**Data Processing / Signal Filtering Choices:** Geophysical data processing is a complex topic with decades of research foundations. Data processing frequently requires choices, including which signals are the “true signal” of interest, versus noise. We wish to highlight how user goals (and therefore the choice of an ultimate “true signal”) may vary by broad community use case. While many data processing techniques in the generation of magnetic anomaly maps are common to the majority of use cases (e.g. correcting for platform noise; measuring baseline solar weather signals; removing data spikes deemed as artifacts; etc), other choices are industry-specific. *(Some choices are even Earth-specific; for example, production of leveled 2D data grids is extremely uncommon in the broader analysis of planetary magnetism, such as for Mars and Jupiter.)*

As an example of industry-specific processing, consider the natural resource industry. In natural resource prospecting, the ultimate user goal is to discover, profile, and monitor key economic deposits. In this sense, the magnetic anomaly field is a tool to discover a different ultimate truth, such as: “*where are the economically relevant deposits?”* Thus, processing focuses intensively on applied analysis, identifying these geologic deposits for users. Data transformations such as Reduced-to-Pole (or Reduction-to-Equator) are essential to that use case, by providing visualizations of the “true” spatial location of an underlying signal. Other kinds of signal processing are also common. Techniques such as low-pass filtering, or more intensive “cultural artifact” removal (human infrastructure-related signal) are utilized to preserve only large-scale, regional signals of geologic origin, at the expense of short-scale signals from other sources.

In contrast, an Operational MagNav user would wish to have a map of the full geomagnetic signal and its uncertainty, exactly as it would appear to a sensor in the field. Continuing with the examples discussed above, maps presented in Reduction-to-Pole form or featuring heavy bandpass filtering would no longer match what an operational user would directly measure during navigation. Careful consideration should be taken with respect to choices made at all stages of the data processing pipeline.

**Format:** Dataset users will benefit greatly from being saved in a common, non-proprietary format that is universally ingestible by modern programming languages. We acknowledge that proprietary data formats used by industry-specific software like ArcGIS or Oasis Montaj are broadly common (and useful) for geophysical work. But ultimately, geophysical anomaly map data for Operational MagNav will need to be queried by team/company-specific software (e.g. navigational filters), which may vary in programming languages and be subject to a certification process. We therefore see implementing a universally compatible dataset format as essential. Many common non-proprietary formats exist (e.g. CSV); we do not advocate for any specific format here, simply for a universal format to be chosen by the community. *Supplementary dataset versions in standard proprietary formats may also be beneficial, but we recommend against this option as a primary/only format.*

**Scalable Querying:** As mentioned above, Operational MagNav Data Users will need to sample the baseline anomaly map and its uncertainty estimate with custom internal software (i.e. navigational filters). Thus, the map will need to be “queryable” by the community. What does this mean? In

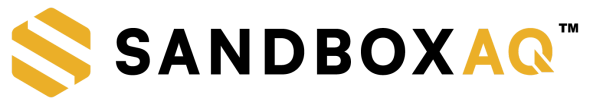

Operational MagNav use cases, navigational filters require quantitative predictions of the anomaly field in full 3D space – where data is available for all latitudes, longitudes, and altitudes within a specific region of validity. It is understood that the Downward Continuation challenge causes mathematical instabilities when maps are projected to lower altitudes, which can increase overall map uncertainty in those regions; this can be dealt with by prescribing a cutoff altitude or bounds of validity for where a map may be used. But broadly speaking, MagNav users will need to easily generate full 3D predictions from input geophysical data (i.e. query the map dataset). This contrasts with other common geophysical use cases, where a 2D view or grid of the geomagnetic field may be sufficient to guide analysis or decision-making.

How can map data be provided in a format that is easily interpolated? The previously discussed properties of minimum **data fields, dataset cohesion, and consistent formatting** will facilitate this, but are not sufficient. Compatibility with common interpolation schemes is also essential. For example, Fourier methods require regular, evenly spaced grids with no holes/data gaps. In contrast, most existing local survey datasets are extremely irregular in shape (due to topographical or scientific requirements), and major global models like EMAG2v3 (Meyer et al., 2017), NAMAM (Bankey et al., 2002), and NAMAM NURE (Ravat et al, 2009) have large, irregular data gaps. When multiple grids at different altitudes are available for a given location (e.g. the Edwards Air Force Base magnetic anomaly map), they are rarely perfectly aligned in latitude/longitude, requiring substantial user massaging before use with those techniques.

An alternative option is to provide a model-based data product. This pathway is more complex, but solves the user's need for ease of interpolation, while providing a consistent solution. In this option, we recommend providing the user with both a model / streamlined API source code, as well as the raw coefficients themselves; see discussion in following paragraphs. Many model options are available, including physical representations such as equivalent sources – dipoles (Mayhew, 1982; Langel & Thoring, 1982), monopoles (O'Brien and Parker, 1994), or pixel variations (Moore et al., 2017), etc; or spectral methods like wavelets (Holschneider et al., 2003), harmonic splines (Shure et al., 1982), Slepians (Simons et al., 2006), or spherical harmonics (e.g. WMM, Chulliat et al. 2020; IGRF, Alken et al., 2021).

In addition to being queryable at a basic level,the querying process and data storage should also be scalable. As a company currently performing MagNav test flights across multiple countries, we are deeply aware of future scaling needs. Specifically, Magnetic field calculations can be computationally intensive, and require large amounts of memory. Sometimes, even the basis functions themselves are not always scalable – for example, popular spherical harmonics packages often encounter computational underflow errors around degree ~2800 (Wieczorek and Meschede, 2018). *(Spherical harmonic degree 2800 corresponds to a spatial wavelength of approximately 14 kilometers, which is likely insufficient for representing crustal magnetic fields at low altitude with high precision).*

Operational MagNav systems may have significant limitations on SWaP-C, so creating a flexible dataset and/or API will be essential. We recommend that the final dataset form be easily divisible by the users into customized smaller sections. For raw dataset-based approaches, this could mean the ability to easily save data into smaller file sizes (a cohesive spatial organization within the file will facilitate this). For API-based implementations, we recommend providing users the ability to apply the model calculations to fully customizable subregions as well as providing the raw source files/model coefficients used by the API. *For example, a scalable equivalent source method API could allow for calculation across a user-specified subset of equivalent sources, and not always require loading the full global source set (since sources far from the user's location will have vanishingly small contributions to the total field). Such an implementation would ease the computational burden on Operational MagNav systems, which may not have sufficient RAM to store the full global model simultaneously in sufficiently high spatial resolution.*

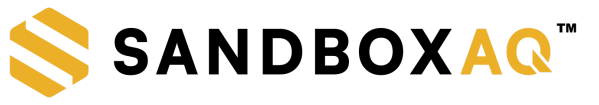

**Documentation & Updates:** We strongly recommend that the datasets be well-documented, with adequate metadata for users to understand the processing and assumptions that have been applied. We also advocate for establishing a process for routine, regular model updates. This has been widely successful for the World Magnetic Model (WMM), another geophysical model routinely used in Navigation.

### 3.3 Magnetic Navigation R&D

The MagNav R&D emphasizes 1) access to original magnetic anomaly survey data for algorithm testing purposes, and 2) ability to understand the data at a deeper level, in order to plan MagNav system test flights. In particular, we suggest the following properties to emphasize. We describe them in detail below, and summarize our recommendations in Table 2.

**Table 2. Priorities for Geophysical Datasets – MagNav R&D Use Case**

| Dataset Property | Recommendation |
|---|---|
| Dataset Cohesion | Original surveys. All data versions – raw flightline data (when available), and any processed/derivative versions |
| Coordinate System | Data presented both in:<br>● Original coordinate system<br>● a Uniform coordinate System *(Standardized & Known across all datasets)* |
| Geophysical Data Fields | Latitude, Longitude, Altitude, Magnetic Anomaly Field, and any additional metadata recorded (solar weather, etc) that was used in processing |
| Magnetic Field Uncertainty Estimates | Localized, queryable uncertainty estimates<br>*(May be provided as separate dataset, or as an additional entry in the above)* |
| Spatial Resolution | Original resolution (maximum resolution; *non-uniform OK)* |
| Altitude | All altitudes that were recorded in original data |
| File Format | Both:<br>● Non-proprietary data format |
| | ● Industry-standard software compatible format. |
| Documentation | Clear documentation describing processing applied to the dataset. Include original surveyor reports, when available |
| Updates | Systematic cadence of dataset updates |
| Searchable Database | Global database of flight datasets, searchable:<br>● Across all available dataset properties *(flight line spacing, size, altitude, year, field uncertainty, etc)*<br>● With an efficient interface, or standard methods<br><br>Ideally, with a searchable GUI option for dataset selection as well, displaying dataset boundaries geographically. |

**Original Survey Data & Documentation:** Central to the MagNav R&D Use Case is access to the original, full-resolution magnetic anomaly data. Whereas Operational MagNav Users will primarily require a unified, cohesive merged dataset, community R&D use cases seek access to data products from earlier in the map making pipeline. This may include regional merged maps created prior to a global merge, as well as the original source magnetic anomaly surveys (both processed/merged grids, and raw flight lines when possible).

Including data products from earlier in the processing pipeline enables MagNav researchers in industry and academia to improve algorithm designs, and conduct critical flight tests to accelerate the technology. For example, having local source grids to compare with the global operational dataset would enable the community to optimize robustness to grid merging choices / artifacts. We note that a global database of survey datasets (including raw flight lines) would also empower the broader geophysics scientific community.

We note that databases for raw original source data and their documentation already exist for many areas of the world. For example, the USGS maintains records for local continental surveys across the United States as well as the NAMAM /NAMAM NURE regional models (Bankey et al., 2002). Recent USGS EarthMRI datasets are available through a separate USGS database. Marine trackline data is available through the NOAA-NCEI. However, we emphasize that while these datasets are available for some regions of the world, they are not always easy to use (see next item) and do not always include original aeromagnetic flight lines.

Most of these datasets already include excellent documentation via metadata or original survey reports; we emphasize that this documentation is useful to the MagNav R&D use case as well. For example, knowing which datasets have had solar weather removed, and which core field model was used for magnetic anomaly calculation would be especially useful.

**Uniform Formatting (Coordinate system, data fields, file format):** Uniform formatting will dramatically increase the usability of magnetic anomaly datasets for MagNav R&D. Specifically, we recommend:

- *A single standardized coordinate system* (e.g. WGS-84 or AGL)

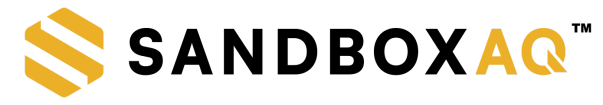

- *Uniform data fields (type, ordering, etc) - Lat, Lon, Alt, Mag Anomaly, uncertainty*
- *Two consistent file formats:*
    - *Standardized, non-proprietary file format*
    - *Industry-software compatible file format*

The rationale for these recommendations is the same as for the Operational MagNav Use Case. But since R&D users will often be comparing multiple datasets, standardization gains increasing emphasis.

**The main exception is that for the MagNav R&D use case, it will also be beneficial to provide data in both a non-proprietary file format, and a file format compatible with standard industry software for geophysical data manipulation** (e.g. ESRI ArcGIS, Seequent Oasis Montaj, etc). This is because users may need to perform custom merging or processing of surveys datasets, which is strongly enabled by these industry tools. Users are more likely to be viewing local survey data (compared to global data for the Operational MagNav use case).

In general, we have found the lack of standardization across existing magnetic anomaly datasets to be a significant barrier to rapid adoption. Currently, data users must custom-write individual data ingest pipelines for each local survey dataset separately. This includes detailed analysis of metadata, to uncover which datum was used for each dataset (occasionally requiring re-reading extensive survey reports for buried information). By providing raw data in a standardized, non-proprietary format, and a single commonly used coordinate system, the efficiency of community innovation can be significantly increased.

**Searchable Database**: An essential need for MagNav R&D is a searchable database for existing survey data. We recommend indexing by key search parameters for MagNav such as *flight line spacing, survey size, altitude, year, and overall uncertainty.* This would allow users to easily find the data that corresponds to their R&D goals, as well as plan flight campaigns based on data availability.

While multiple databases of survey data exist (USGS, NOAA, etc), they are not all fully searchable across key properties relevant to MagNav R&D. For example, the EarthMRI database is searchable only by overall dataset type (Geophysics, Geology, Lidar, etc), Project Name, Affiliation, Team Name, or Year (USGS). We have also found GUI-based interfaces to be extremely helpful for planning field campaigns, as they display dataset boundaries and overlaps geographically.

### 3.3.1 Vector Magnetic Anomaly Data.

While this White Paper emphasizes scalar magnetic anomaly data as the current operational baseline for MagNav, vector anomaly measurements represent an important area of ongoing research. Vector data can provide additional directional information that may improve observability, robustness to interference, and performance in certain environments. However, the collection, processing, and standardization of vector anomaly datasets remain less mature than their scalar counterparts. As such, we view vector anomaly data as a priority for MagNav R&D test ranges and experimental campaigns, rather than an immediate requirement for global operational datasets.

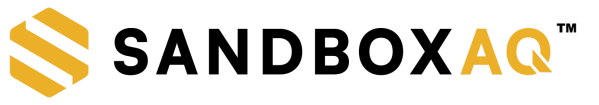

Early inclusion of vector datasets in controlled test environments will allow the community to assess their value systematically and inform future standards evolution.

## 4. Core Field Modeling: Expansion of the WMM to deg 13

Finally, we seek to highlight a supplementary but essential model for magnetic anomaly calculations: a standardized model for Earth's core field. Magnetic anomaly datasets are typically calculated by removing a model of Earth's core field to spherical harmonic degree and order 13. However, the currently required geophysical core field model in Dept. of Defense use cases, the WMM, only extends to degree 12 (Chulliat et al., 2020). This means that if the deg-12 WMM were required to be used for MagNav, significant magnetic signals would be unaccounted for, degrading overall navigational performance.

**We recommend that future versions of the World Magnetic Model (or similar data product) extend coefficient models to degree and order 13, to ensure consistency with how magnetic anomalies are calculated scientifically**. We also recommend that the Operational MagNav geophysical dataset(s) be uniformly processed, where a single consistent core field model is removed from all underlying source data. This will ensure there are no changes in DC baseline between regions if different core field models were removed from each *(we believe this is a likely explanation for some artifacts we have encountered in existing public datasets)*.

## 5. Discussion: MagNav Test Ranges

In the previous Sections, we recommend high-level priority areas for MagNav geospatial data requirements and products. However, we empathize that this will not be an instantaneous undertaking. Specific geographic areas and dataset types will need to be prioritized, in order for data to be made available more quickly.

**As a MagNav geophysical data user, we believe a top priority will be the organization of MagNav "Test Ranges".** In these regions, multiple geomagnetic anomaly datasets would be made available and concentrated in a single location. **Example test ranges are currently being organized – an endeavor that is much welcome by the navigation community.** See for example, the AngelWings reference dataset (Brinkley et al., 2024).

**We strongly agree that Test Ranges will be essential to both MagNav validation (Operational MagNav), as well as ongoing MagNav R&D.** They present an excellent opportunity to roll out side-by-side datasets for both use cases – geographically regioned snippets of an Operational MagNav magnetic anomaly dataset, as well as the underlying source survey data. This will allow for rapid adoption and enablement of MagNav, as well as continuous feedback between the community for dataset development.

**We suggest that the location selection for potential Test Ranges factor in multiple geographic considerations.** To begin with, each MagNav modality will have unique needs from a test range. For example, in Aerial MagNav, dataset coverage over different terrain types and altitudes will be essential. For Maritime MagNav, sea level maps covering different ocean depths will be desired.

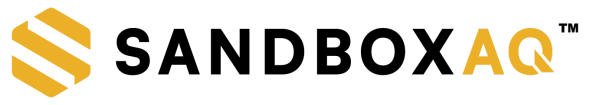

Data over transitions, such as between land and sea will also be helpful (e.g. AngelWings; Brinkley et al., 2024). Finally, both aerial and maritime modalities will eventually desire access to a test region for longer-range flight (e.g. a 4-6 hour long-haul flight or voyage; essential to many Commercial and Defense navigation needs), either as a continuous dataset over the continental USA, and/or as transoceanic testing range. We see a high-resolution, long-haul test range as an outstanding need.

**Multiple test ranges should be made available in multiple different geographic regions, to facilitate testing by industry and research institutions across the country.** Aerial technology testing is expensive, and often billed by the hour. Reservation windows aboard designated test aircraft are also extremely limited, and may require booking months in advance. In such an environment, flying 6 hours to bring the desired test aircraft to a MagNav test range on the opposite coast could be cost- or logistically-prohibitive. By creating test ranges in multiple geographic regions, the speed of community technology development could be substantially improved.

**Another important geographic factor for Test Range locations is local Air Space Regulations.** In our experience, regions with heavy air traffic usage can impose substantial limitations on MagNav flight scheduling. For example, low altitude flight is essential for many PNT needs such as airport approaches or specialized evasive maneuvers. But low-altitude test flights can be more challenging to schedule, especially over populated continental areas. Only a limited number of Low Level routes exist that are pre-approved by Air Traffic Control / the Federal Aviation Administration (these are typically accompanied by ratings for either visual or instrument-based flight). Having Test Ranges that intersect with specialized routes such as these is significantly advantageous for testing.

MagNav test ranges that explicitly support unmanned aerial system (UAS) operations would provide substantial benefits to the community. UAS platforms enable cost-effective, repeatable, and low-altitude testing—regimes that are particularly valuable for characterizing magnetic anomaly sensitivity and map fidelity near the surface. Under current U.S. regulatory frameworks, UAS flights are typically permitted up to approximately 400 ft AGL in Class G airspace, while operations beyond this envelope require formal waivers or designated test environments. Considering UAS-accessible airspace in the design and selection of MagNav test ranges would therefore reduce regulatory friction, expand platform diversity, and support a broader range of users and technology readiness levels.

Additionally,Tolles-Lawson maneuvers (a common Figure of Merit for MagNav; Tolles and Lawson, 1950) can also require a substantial geographic area for completion depending on the platform's speed and turn radius. While not always needed for Operational MagNav, they are still a common maneuver in MagNav R&D. However, we have found that these larger spatial maneuvers can be difficult to organize over densely populated areas, due to air traffic controls. Thus, flight density / underlying air traffic regulations can play an important role in the effectiveness of a MagNav test range. *We note that regions with higher air traffic are more likely to be affiliated with major international airports or military bases, and so there may be some tradeoffs between accessibility of the Testing Range (i.e. ease of personnel travel to/from the test range) and ease of use for MagNav testing.*

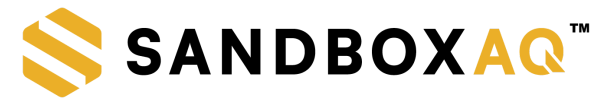

Military Use areas also naturally pose restrictions on test flights. For example, the magnetic anomaly map of Edwards Air Force Base is commonly cited as a pristine map for MagNav flight tests (e.g. Canciani, 2021). But due to its location over an active military base, we have found large portions of the map to be challenging to access for broader commercial aircraft or industry test flights. On the other hand, we note that having test ranges in areas restricted for military use also provides distinct advantages; certain types of PNT testing (e.g. hypersonic testing) likely may only be performed in these areas. **We recommend test ranges consider thoughtful enablement of both commercial and defense MagNav testing needs.**

**We also recommend MagNav test ranges include uncommon or unusual data types such as underwater or vector anomaly data.** While magnetic anomaly data (of varying quality/spatial resolution) is readily available for sea level or higher, the same cannot be said about subsurface data. In order to enable MagNav testing for underwater modalities, access to data in a testing range will be important. By the same logic, ground-level magnetic anomaly maps, and anomaly maps of urban areas will also be essential for further modalities. Finally, providing Test Ranges for promising MagNav data types, like vector magnetic anomaly data, will facilitate future MagNav R&D.

Beyond enabling validation and experimentation, MagNav test ranges provide a critical mechanism for informing formal geophysical data requirements. Systematic comparison of map products, uncertainty models, altitude performance, and processing choices within shared test environments can generate empirical evidence to support standards development. We recommend that insights derived from test range activities be explicitly captured and fed back into future requirements processes, ensuring that standards evolve based on demonstrated operational performance rather than purely theoretical considerations.

Beyond the availability of testing ranges, we also highlight additional considerations for the prioritization of rollout Operational MagNav data globally. Specifically, areas with present and ongoing GPS/GNSS outages will be a key priority, as is ensuring robust PNT for the operational theatres of tomorrow. **We advocate for sustained collaboration with the Armed Services and Intelligence communities, to identify priority geographic regions for Operational MagNav Datasets.**

## 6. Conclusions

MagNav is an essential technology for alt-PNT strategy, to enable global navigation for commercial and defense use cases alike. Standardized data products for MagNav will enable rapid technology deployment for today (Operational MagNav), as well as future development of even more advanced MagNav algorithms for tomorrow (MagNav R&D). Active support from the Armed Services –as well as the Intelligence and broader Industry communities – will be essential to the development of MagNav dataset requirements. We hope that this White Paper is just the beginning of that discussion.

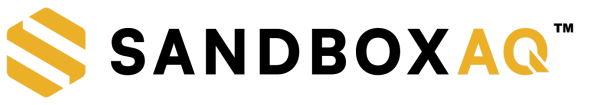

### 6.1 Call to Action.

We encourage government agencies, data providers, industry, and the research community to use this White Paper as a starting point for coordinated requirements development. Near-term actions could include convening a regularly convened, focused working group on MagNav geophysical data standards, prioritizing the establishment of shared test ranges, and initiating pilot efforts to develop uncertainty-aware, queryable magnetic anomaly datasets. Early alignment across stakeholders will accelerate the transition of MagNav from demonstrated capability to operationally assured navigation infrastructure.

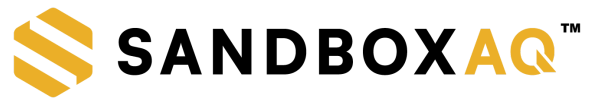

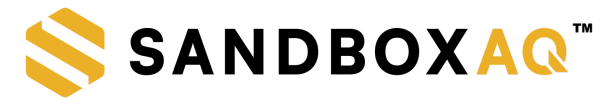

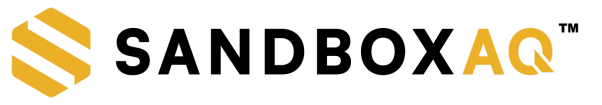

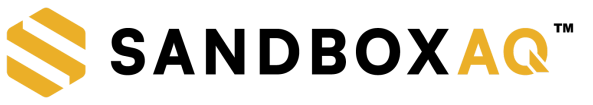